\documentclass[10pt,letterpaper]{article}
\usepackage{multirow}

\usepackage[top=0.85in,left=2.75in,footskip=0.75in]{geometry}

\usepackage{amsmath,amssymb}

\usepackage{changepage}

\usepackage[utf8x]{inputenc}

\usepackage{textcomp,marvosym}

\usepackage{cite}

\usepackage{nameref,hyperref}

\usepackage[right]{lineno}

\usepackage{microtype}
\DisableLigatures[f]{encoding = *, family = * }

\usepackage[table]{xcolor}

\usepackage{array}

\newcolumntype{+}{!{\vrule width 2pt}}

\newlength\savedwidth

\raggedright
\usepackage[aboveskip=1pt,labelfont=bf,labelsep=period,justification=raggedright,singlelinecheck=off]{caption}

\makeatletter
\renewcommand{\@biblabel}[1]{\quad#1.}
\makeatother

\usepackage{lastpage,fancyhdr,graphicx}
\usepackage{epstopdf}
\fancyheadoffset[L]{2.25in}
\fancyfootoffset[L]{2.25in}
\usepackage{xcolor}
\newcommand{\hl}[1]{\textcolor{black}{#1}}
\begin{document}
\vspace*{0.2in}

\begin{flushleft}
{\Large
\textbf\newline{Friendly-rivalry solution to the iterated $n$-person public-goods game} % Please use "sentence case" for title and headings (capitalize only the first word in a title (or heading), the first word in a subtitle (or subheading), and any proper nouns).
}
\newline
\\
Yohsuke Murase\textsuperscript{1} and
Seung Ki Baek\textsuperscript{2*}
\\
\bigskip
\textbf{1} \hl{RIKEN Center for Computational Science, Kobe, Japan}
\\
\textbf{2} \hl{Department of Physics, Pukyong National University, Busan,
Korea}
\\
\bigskip

% Insert additional author notes using the symbols described below. Insert symbol callouts after author names as necessary.
%
% Remove or comment out the author notes below if they aren't used.
%
% Primary Equal Contribution Note
%\Yinyang These authors contributed equally to this work.

% Additional Equal Contribution Note
% Also use this double-dagger symbol for special authorship notes, such as senior authorship.
%\ddag These authors also contributed equally to this work.

%\textcurrency Current Address: Dept/Program/Center, Institution Name, City, State, Country % change symbol to "\textcurrency a" if more than one current address note
% \textcurrency b Insert second current address
% \textcurrency c Insert third current address


% Group/Consortium Author Note
%\textpilcrow Membership list can be found in the Acknowledgments section.

% Use the asterisk to denote corresponding authorship and provide email address in note below.
* seungki@pknu.ac.kr

\end{flushleft}
\section*{Abstract}
Repeated interaction promotes cooperation among rational individuals under the shadow of future, but it is hard to maintain cooperation when a large number of error-prone individuals are involved. One way to construct a cooperative Nash equilibrium is to find a `friendly\hl{-}rivalry' strategy, which aims at full cooperation but never allows the co-players to be better off. Recently it has been shown that for the iterated Prisoner's Dilemma in the presence of error, a friendly rival can be designed with the following five rules: Cooperate if everyone did, accept punishment for your own mistake, punish defection, recover cooperation if you find a chance, and defect in all the other circumstances. In this work, we construct such a friendly-rivalry strategy for the iterated $n$-person public-goods game by generalizing those five rules. The resulting strategy makes a decision with referring to the previous $m=2n-1$ rounds. A friendly-rivalry strategy for $n=2$ inherently has evolutionary robustness in the sense that no mutant strategy has higher fixation probability in this population than that of a neutral mutant. Our evolutionary simulation indeed shows excellent performance of the proposed strategy in a broad range of environmental conditions when $n= 2$ and $3$.

% Please keep the Author Summary between 150 and 200 words
% Use first person. PLOS ONE authors please skip this step.
% Author Summary not valid for PLOS ONE submissions.
\section*{Author summary}
How to maintain cooperation among a number of self-interested
individuals is a difficult problem, especially if they can sometimes commit
error. In this work, we propose a strategy for the iterated $n$-person
public-goods game based on the following five rules: Cooperate if everyone did,
accept punishment for your own mistake, punish others' defection, recover
cooperation if you find a chance, and defect in all the other circumstances.
These rules are not far from actual human behavior, and
the resulting strategy guarantees three advantages: First, if everyone uses
it, full cooperation is recovered even if error occurs with small probability.
Second, the player of this strategy always never obtains a lower long-term
payoff than any of the co-players. Third, if the co-players are unconditional
cooperators, it obtains a strictly higher long-term payoff than theirs.
Therefore, if everyone uses this strategy, no one has a reason to change it.
Furthermore, our simulation shows that this strategy will become highly
abundant over long time scales due to its robustness against the invasion of
other strategies. In this sense, the repeated social dilemma is solved for an
arbitrary number of players.
%\linenumbers

% Use "Eq" instead of "Equation" for equation citations.
\section*{Introduction}

The success of {\it Homo sapiens} can be attributed to its ability to organize
collective action toward a common goal among a group of genetically unrelated
individuals~\cite{nowak2011supercooperators}, and this ability is becoming more
and more important as the world is getting close to each other.
Researchers have identified several mechanisms to promote cooperation in terms
of evolutionary game theory~\cite{nowak2006five,sigmund2010calculus}. For
example, the folk theorem holds that repeated interaction can establish
cooperation through reciprocal strategies, and this mechanism is called direct
reciprocity~\cite{fudenberg1991game}.
Yet, how to resolve a conflict between individual and collective interests is
a hard problem, especially when a large number of players are involved and they
are prone to
error~\cite{molander1985optimal,boyd1988evolution,gintis2006behavioral}, because
an individual player has very limited control over co-players.

In this respect, the discovery of the zero-determinant (ZD) strategies in the
iterated prisoner's dilemma (PD) has been
deemed counter-intuitive because a ZD-strategic player can unilaterally fix the
co-player's long-term payoff or enforce a linear relationship between their
long-term payoffs~\cite{press2012iterated}.
For instance, one can design an \emph{extortionate} ZD strategy, with which
the player's long-term payoff will increase by $\chi \ge 1$ whenever the
co-player's does by one unit payoff. Another counter-intuitive aspect of the
ZD strategy is that it is a memory-one strategy referring only to the previous
round, so that such a simple strategy can perfectly constrain the co-player's
long-term payoff regardless of the co-player's strategic complexity.
Of course, the excellent performance in a one-to-one match does not
necessarily mean evolutionary success: It is difficult for an extortionate
strategy to proliferate in a population because, as its fraction increases,
two extortionate players are more likely to meet and keep defecting against each
other~\cite{hilbe2013evolution,hilbe2013adaptive,stewart2013extortion,adami2013evolutionary}.
For this reason, especially in a large population, selection tends to favor a
\emph{generous} ZD strategy whose long-term payoff does not exceed the
co-player's~\cite{stewart2013extortion}. A generous ZD strategy
does not aim at winning a match, but
it is efficient by forming mutual cooperation when they meet each other.

The important point in this line of thought is that a player's strategy
can unilaterally impose constraints on the co-player's long-term payoff, so that
we can now characterize strategies according to the constraints that they
impose.
One such meaningful characterization scheme is to ask if a strategy works as a
`partner' or as a `rival'~\cite{hilbe2015partners,hilbe2018partners}: By
`partner', we mean that the strategy seeks for mutual cooperation, but that it
will make the co-player's payoff less than its own if the co-player defects from
it. It has also been called `good'~\cite{akin2015you,akin2016iterated},
and the generous ZD strategies
can be understood as an intersection between the ZD and partner
strategies~\cite{stewart2013extortion}. On the other hand, a rival strategy
always makes its long-term payoff higher than or equal to the co-player's, so it
has been called `unbeatable'~\cite{duersch2012unbeatable},
`competitive'~\cite{hilbe2015partners}, or
`defensible'~\cite{yi2017combination,murase2018seven}. A trivial example of a
rival strategy is unconditional defection (AllD), and an extortionate ZD
strategy also falls into this class.
Most of well-known strategies in the iterated PD game are classified either as a
partner or as a rival~\cite{hilbe2018partners}. However, which class is more
favored by selection depends on environmental conditions such as the population
size and the benefit-to-cost ratio of cooperation:
If the population is small and cooperation is costly, it is better off to play a
rival strategy than to play a partner strategy, and vice
versa~\cite{stewart2013extortion,stewart2014collapse,hilbe2018partners}.
\hl{In the iterated PD game,}
if a single strategy acts as a partner and a rival simultaneously, it has
important implications in evolutionary dynamics because it possesses
\emph{evolutionary robustness} regardless of the environmental
conditions~\cite{murase2020five}, in
the sense that no mutant strategy can invade a population of this strategy
with greater fixation probability than that of neutral
drift~\cite{stewart2013extortion,stewart2014collapse,stewart2016small}.
To indicate the partner-rival duality, such a strategy will be called a
`friendly rival'~\cite{murase2020five}. Tit-for-tat (TFT), a special ZD strategy
having $\chi=1$, is a friendly rival in an error-free
environment~\cite{hilbe2018partners}, but a friendly rival generally requires a
far more complicated structure in the presence of error. So far, the existence
of friendly-rivalry strategies has been reported by a brute-force enumeration
method in the iterated PD
game~\cite{baek2008intelligent,yi2017combination,murase2020five} and the
three-person public-goods (PG) game~\cite{murase2018seven}. However,
it is not straightforward to extend these findings to the general $n$-person PG
game. For example, a naive extension of a solution in the iterated PD game
fails to solve the three-person PG game because the third player cannot tell
if one of the co-players is correcting the other's error with good intent or
just carrying out a malicious attack~\cite{murase2018seven}.
To resolve the ambiguity, a strategic decision must be based on more
information of the past interactions: In fact, if a player refers to the
previous $m$ rounds to choose an action in the $n$-person PG game, we can show
that $m$ must be greater than or equal to $n$ as a necessary condition to be a
friendly rival~\cite{murase2018seven}. Unfortunately, the existing brute-force
approach then becomes simply unfeasible because the number of possible
strategies expands super-exponentially as $2^{2^{mn}}$. For example, in the
three-person game ($n=3$), it means that we have to enumerate $2^{512} \sim
10^{154}$ possibilities to find an answer. Although the symmetry among
co-players reduces this number down to $2^{288} \sim 10^{86}$, it is still
comparable to the estimated number of protons in the universe.

\begin{figure}
\begin{center}
\includegraphics[width=0.6\textwidth]{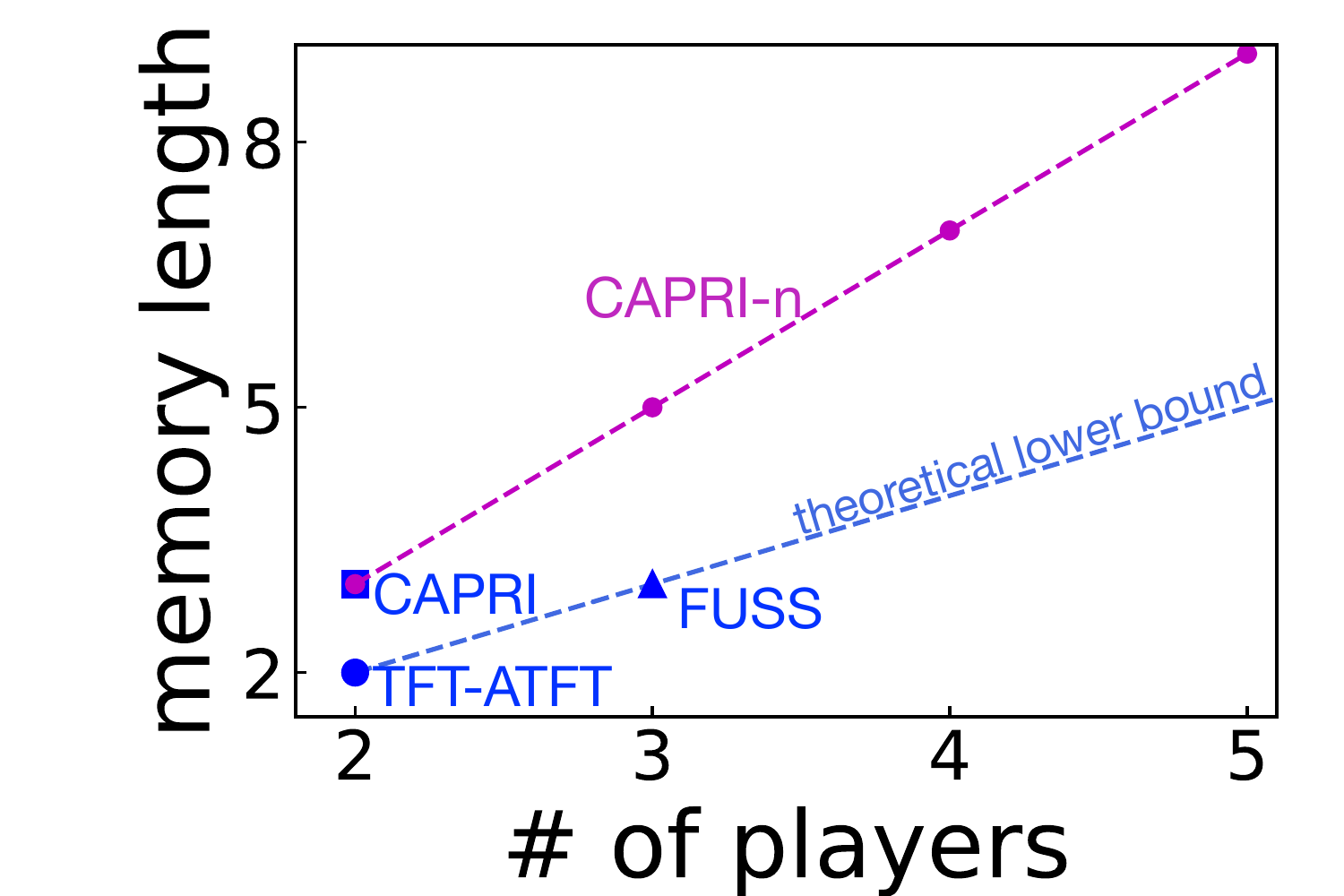}
\end{center}
\caption{Memory length $m$ required for each of currently known friendly-rivalry
strategies in the
$n$-person PG game~\cite{yi2017combination,murase2018seven,murase2020five}.
The dashed blue line depicts a theoretical lower bound $m = n$ for friendly
rivalry~\cite{murase2018seven}, and the strategy proposed in this work, called
CAPRI-$n$, has $m=2n-1$.
}
\label{fig:m_n}
\end{figure}

In this work, by using an alternative method to generalize behavioral
patterns of a friendly rival for the iterated PD game~\cite{murase2020five},
we construct a friendly-rivalry strategy for the $n$-person PG game.
This approach makes use of the fact that it greatly lessens the computational
burden if we only check whether a given strategy qualifies as a friendly rival.
The required memory length of our strategy is $m = 2n-1$, which
satisfies the necessary condition $m \geq n$ as shown in \hl{Fig}~\ref{fig:m_n}.
We will also numerically confirm that it shows excellent performance in
evolutionary dynamics. In this way, this work modifies and
generalizes our previous finding of $n=2$ and $m=3$, i.e., a memory-three
friendly-rivalry strategy for the iterated PD game~\cite{murase2020five}.

\section*{Materials and methods}

\hl{In this section, we define the game and construct a friendly-rivalry strategy by
reasoning. See \nameref{S1_Table} for a summary of mathematical symbols.}

\subsection*{Public-goods game}
Let us consider the $n$-person public-goods (PG) game, in which a player may
choose either cooperation ($c$), by contributing a token to a public pool, or
defection ($d$), by refusing it. Let the number of cooperators be denoted as
$n_c$. The $n_c$ tokens in the public pool are multiplied by a factor of
$\rho$, where $1<\rho<n$, and then equally redistributed to the $n$ players. We
assume that the tokens are infinitely divisible. A player's payoff is thus
given as
\begin{equation}
  \begin{cases}
    \frac{\rho n_c}{n} & \text{when the player chooses $c$,} \\
    1 + \frac{\rho n_c}{n} & \text{when the player chooses $d$.} \\
  \end{cases}
\label{eq:payoffs}
\end{equation}
Clearly, it is always better off to choose $d$ regardless of $n_c$, so full
defection is the only Nash equilibrium of this one-shot game. In this study,
this game will be repeated indefinitely with no discounting factor to
facilitate direct reciprocity. Every player can choose an action between $c$ and
$d$ by referring to the previous $m$ rounds. At the same time, a player can make
implementation error, e.g., by choosing $d$ while intending $c$ and vice versa,
with small probability $\epsilon \ll 1$.

\subsection*{Axiomatic approach}
% Let us consider a strategy profile $\mathcal{P} =\{\Sigma_1, \Sigma_2, \ldots,
% \Sigma_n\}$ of $n$ players.
The long-term payoff of player $X$ is defined as
\begin{equation} \Pi_{X} \equiv \lim_{T\to\infty}
\frac{1}{T} \sum_{t=0}^{T-1} \pi_{X}^{(t)},
\end{equation}
where $\pi_{X}^{(t)}$ is player $X$'s instantaneous payoff in round $t$.
If $\epsilon>0$,
the Markovian dynamics of strategic interaction for a given strategy
profile converges to a unique stationary distribution, from which
$\Pi_X$ can readily be calculated~\cite{nowak1990stochastic,nowak1995automata}.
In terms of the players' long-term payoffs, we wish to propose the following
three criteria that a successful strategy $\Omega$ should
satisfy~\cite{yi2017combination,murase2018seven,murase2019automata,murase2020five}.
\begin{enumerate}
  \item
    Efficiency: Mutual cooperation must be achieved with probability one as
    $\epsilon \to 0$ if all the players have adopted $\Omega$.
    In other words, this criterion requires $\lim_{\epsilon\to 0^+}\Pi_{X} =
    \rho$ when the strategy profile $\mathcal{P} = %\mathcal{P}_{\Omega} \equiv
    \{ \Omega, \Omega,\ldots, \Omega \}$.
  \item
    Defensibility: It must be guaranteed that none of the co-players can
    obtain higher long-term payoffs against $\Omega$ regardless of their
    strategies and initial states when $\epsilon=0$. It implies that
    $\lim_{\epsilon \to 0^+} \left( \Pi_X - \Pi_Y \right) \ge 0$, where player
    $X$ is using strategy $\Omega$ and $Y$ denotes any possible co-player of
    $X$.
  \item
    Distinguishability: If $X$ uses $\Omega$ and all the co-players are
    unconditional cooperators (AllC), player $X$ can exploit them to earn a
    strictly higher long-term payoff than theirs. That is,
    $\Pi_{X} > \Pi_{Y}$ when $Y$ is an AllC player.
\end{enumerate}
When a strategy satisfies defensibility and efficiency, the strategy is a
friendly rival. A symmetric strategy profile which consists of a
friendly-rivalry strategy forms a cooperative Nash
equilibrium~\cite{yi2017combination,murase2018seven,murase2020five},
and the proof is straightforward: Assume that everyone initially
uses a friendly-rivalry strategy $\Omega$ with earning $\rho$ per round. If one
player, say, $X$, changes his or her strategy alone, $X$'s payoff will
change to $\Pi_X$, while each of the co-players earns $\Pi_\Omega$.
Defensibility guarantees that $\Pi_X \le
\Pi_\Omega$, and full cooperation is Pareto-optimal, i.e., $(n-1)\Pi_\Omega +
\Pi_X \le n \rho$. Combining these two inequalities, we see that
\begin{equation}
(n-1) \Pi_X + \Pi_X \le (n-1) \Pi_\Omega + \Pi_X \le n\rho,
\end{equation}
which means that $\Pi_X \le \rho$. The player cannot
increase his or her payoff by deviating from $\Omega$ alone.
The third criterion is a requirement
to suppress invasion of AllC due to neutral drift in the evolutionary
context~\cite{imhof2005evolutionary,imhof2007tit,imhof2010stochastic}.
We call a strategy `successful' if it meets all the above three criteria.
Depending on the definition of successfulness, one could choose a different set
of axioms for an alternative characterization~\cite{hilbe2017memory}.

\subsection*{Strategy design}

Let us construct a deterministic strategy with memory length $m = 2n - 1$
and show that the proposed strategy indeed meets all of the above three
criteria.
In the following, we will take an example of three players ($n=3$) who are
called Alice ($A$), Bob ($B$), and Charlie ($C$), respectively, and choose Alice
as a focal player playing this strategy.

Before proceeding,
it is convenient to introduce some notations for the sake of brevity.
The three players' history profile over the previous $m=5$ rounds can be
represented as $h_t=(A_{t-5}A_{t-4}A_{t-3}A_{t-2}A_{t-1};
B_{t-5}B_{t-4}B_{t-3}B_{t-2}B_{t-1}; C_{t-5}C_{t-4}C_{t-3}C_{t-2}C_{t-1})$,
where $A_{\tau}$, $B_{\tau}$, and $C_{\tau}$ denote their respective actions at
round $\tau$. The last round of full cooperation will be denoted by $t^\ast$.
%In addition, we introduce a binary variable $\lambda_X^{(t)}$ which equals
%one if $X_t = d$ and zero otherwise for player $X \in \{A, B, C\}$.
%According to the payoff definition [Eq.~\eqref{eq:payoffs}], Alice's payoff in round
%$t$ can be rewritten as
%\begin{equation}
%\pi_A^{(t)} = \rho \left[ 1- \frac{1}{n}\sum_{X} \lambda_X^{(t)}
%\right] - \left[1-\lambda_A^{(t)} \right],
%\end{equation}
%which has linear dependence on $\lambda_X^{(t)}$ for every $X$.
%This linearity
According to the payoff definition [Eq.~\eqref{eq:payoffs}], we can fully
determine Alice's cumulative payoff over a given period, $\sum_t \pi_A^{(t)}$,
just by counting how many times each of the players has defected during the
period. This is due to the linearity of operations acting on the number of
tokens: The tokens contributed to the public pool are multiplied by a constant
factor $\rho$ and equally distributed to all the players, and Alice saves a
token every time she defects.
For example, if all the players have defected the same number of
times, their payoffs must be the same irrespective of the exact history.
We thus introduce $\Delta_A^{\tau_1, \tau_2}$ to denote Alice's number of defections in
$[\tau_1, \tau_2]$. Likewise, we can define $\Delta_B^{\tau_1,
\tau_2}$ for Bob and $\Delta_C^{\tau_1, \tau_2}$ for Charlie.
We also define $N_d$ as the maximum difference among
the players in numbers of defections over the previous $m$ rounds:
\begin{equation}\label{eq:nd}
N_d \equiv \max_{i, j \in \{A,B,C\}} \left|
\Delta_i^{t-m, t-1} - \Delta_j^{t-m, t-1} \right|.
\end{equation}

With these notations, we can now design a successful strategy satisfying all
the three criteria simultaneously. To this end,
we divide the set of history profiles into three
mutually exclusive cases: The first case is that full cooperation occurred
in the last round ($t^\ast = t-1$).
The second case is that it is not in the last round but still in their
memory ($t-m \le t^\ast < t-1$). The third case is that no player remembers the
last round of full cooperation ($t^\ast < t-m$).
Let us consider these cases one by one, together with adequate rules for
each.

\begin{figure}
\begin{center}
\includegraphics[width=0.9\textwidth]{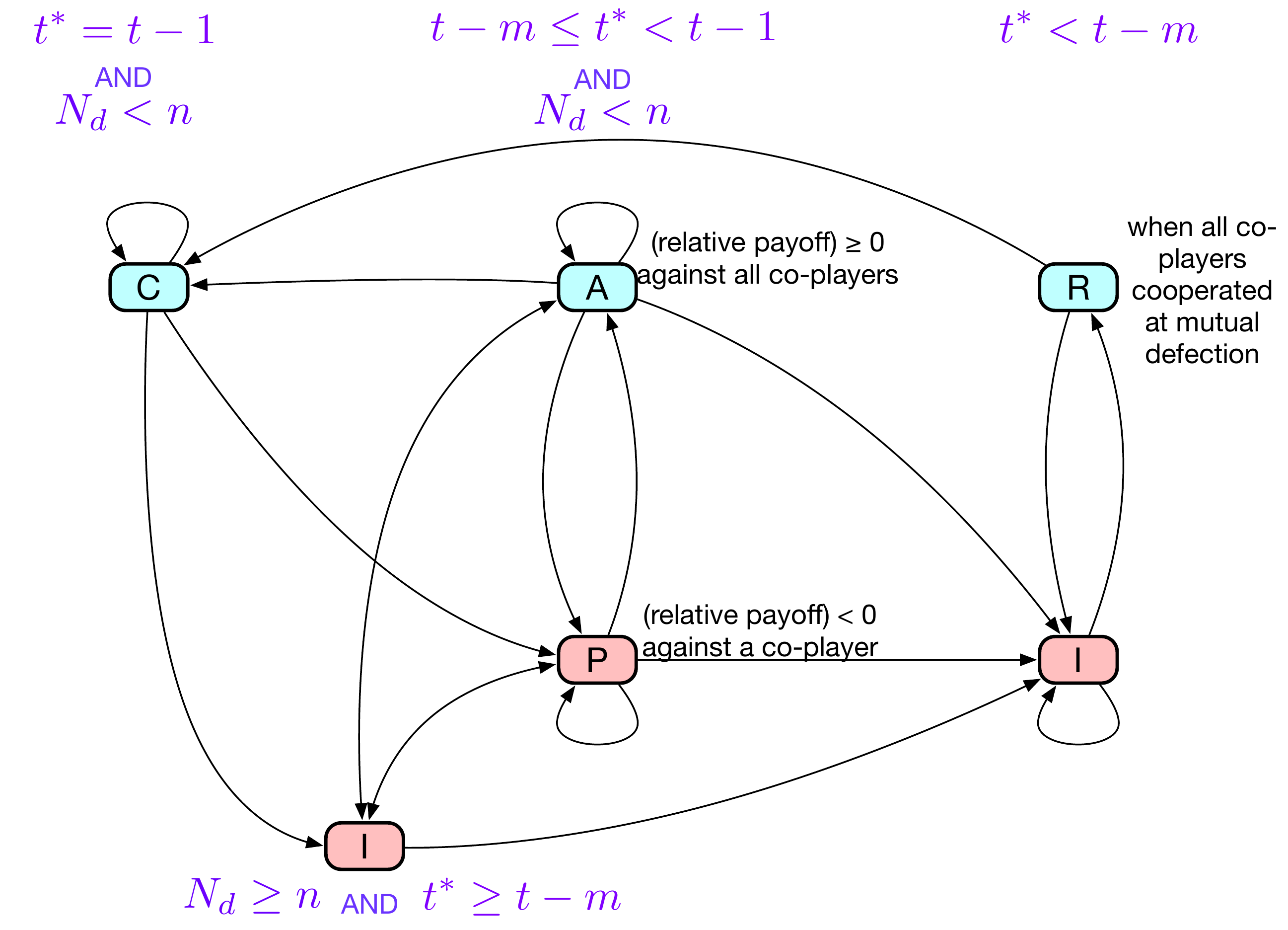}
\end{center}
\caption{
Schematic diagram of the transition between states of CAPRI-$n$. The five rules
of the strategy can be identified with the player's internal
states~\cite{murase2019automata}, each of which is represented as a node in this
diagram.
An exception is state I, which corresponds to two nodes to clarify the following
point: When $t^{\ast} \geq t-m$, the state may have outgoing connections to
\hl{A and P}.
\hl{
When $t^{\ast} < t-m$, on the other hand, the only possible next state is R.
The player has to choose $c$ at a blue node and $d$ at a red node.}
We have omitted error-caused transitions for the sake of simplicity.
}
\label{fig:states_CAPRI_n}
\end{figure}

\begin{enumerate}
    \item $t^\ast = t-1$
    \begin{itemize}
    \item Cooperate: If this is the case, Alice has to choose $c$ under the
    condition that $N_d < n$. For example, the inequality is true for
    $(ccccc;cccdc;ccccc)$, for which $N_d = 1$. On the other hand, it
    is not true for $(cdddc;ccddc;ccccc)$ \hl{because its $N_d$ is
    equal to $n=3$.}
    \end{itemize}
    \item $t-m \le t^\ast < t-1$
    \begin{itemize}
        \item Accept:
        Alice has to accept punishment from the co-players by choosing $c$,
        under the condition that $\Delta_A^{t^\ast, t-1} \ge \Delta_B^{t^\ast,
        t-1}$ and $\Delta_A^{t^\ast, t-1} \ge \Delta_C^{t^\ast, t-1}$ in
        addition to $N_d<n$.
        For example, $c$ will be prescribed to Alice at $(cccdc; ccccd,
        ccccc)$, where we have $t^\ast = t-3$, $\Delta_A^{t^\ast, t-1} =
        \Delta_B^{t^\ast, t-1}=1$, $\Delta_C^{t^\ast, t-1}=0$, and $N_d = 1$,
        which satisfies the above inequalities.
        On the other hand, the condition is not met by $(ccddd; ccddd; ccccc)$
        which gives $N_d = 3$.
        \item Punish: Alice has to punish the co-players by choosing $d$,
        under the condition that $\Delta_A^{t^\ast,t-1}<\Delta_B^{t^\ast,t-1}$
        or $\Delta_A^{t^\ast,t-1}<\Delta_C^{t^\ast,t-1}$ in addition to
        $N_d < n$.
        For example, $d$ is prescribed at $(ccccd; cccdd; ccccc)$
        because $N_d=2$ and Alice has defected fewer times than Bob since
        the last round of full cooperation at $t^{\ast}=t-3$.
    \end{itemize}
    \item $t^\ast < t-m$
    \begin{itemize}
    \item Recover:
    Alice has to recover cooperation by choosing $c$,
    under the condition that all the players except one cooperated in the
    last round. For $n=3$, it means $(ddddd;ddddc;ddddc)$ and its permutations.
    \end{itemize}
    \item In all the other cases, defect.
\end{enumerate}
A strategy of this sort for the $n$-person PG game will be called CAPRI-$n$
after the first letters of the five constitutive rules. Note that these five
rules may be implemented in a number of different ways~\cite{murase2020five},
and we take this way because it provides the most
straightforward way to prove the three criteria. Each of the rules can also be
regarded as the player's internal \emph{state} consisting of multiple history
profiles~\cite{murase2019automata}. For example, Alice can find herself at state
R, the abbreviation for `Recover', when her history profile is $(ddddd;ddddc;ddddc)$,
at which she must choose $c$.
%The other states from C to
%I can be understood in a similar way.
The connection structure of the above five
states is graphically represented in \hl{Fig}~\ref{fig:states_CAPRI_n}, which is helpful
for understanding how defensibility and efficiency are realized as shown below.

\subsubsection*{Defensibility}
Let us begin by checking defensibility.
Our CAPRI-$n$ player Alice cooperates only at states C, A, and R, so the
question is whether she can be forced to visit one of these states repeatedly
with giving a strictly higher payoff to one of her co-players.
If Alice's state
is C, it means that everyone cooperated at $t-1$.
If some of her co-players defect from this full cooperation at $t$,
she will retaliate at $t+1$ with state P, so she suffers from unilateral
defection at most once. Full cooperation is already broken, so it must be only
through state A or R if she comes back to C.
The former case, i.e. the case when she comes back to C via A, means that Alice
has already been compensated for the payoff loss: Otherwise, she would not have
state A.
In the latter case when C is accessed via R, on the other hand, the only possible history profile is $(ddddd;ddddc;ddddc)$
unless she made a mistake, which means the compensation has been done in the last round.
Finally, state A can be accessed from
states P and I, at both of which one cannot exploit Alice who chooses $d$. To
sum up, it is impossible to \hl{see unilateral cooperation} of a CAPRI-$n$ player
repeatedly.

\subsubsection*{Efficiency}
The next criterion is efficiency. Provided that CAPRI-$n$ is employed by
all the players, only full cooperation or full defection can be a
stationary state, and we can verify this statement by checking each possible
case:
\begin{itemize}
\item If $t^\ast=t-1$, everyone \hl{has} to cooperate again as prescribed at state
C, so full cooperation will continue.
\item If $t-m \le t^\ast < t-1$ and
$N_d<n$, some players must be at state A while the others are at
state P. The latter players at P will keep defecting until satisfying
$\Delta_A^{t^\ast, t-1} = \Delta_B^{t^\ast, t-1} = \Delta_C^{t^\ast, t-1}$.
If they make it with keeping $t^\ast \ge t-m$, all of them should choose $c$ as
prescribed at state A, and the resulting mutual cooperation will continue.
If they don't, the situation to everyone reduces to state I, at which they will
defect over and over.
\item The remaining state is R, but it is always transient.
\end{itemize}
To judge efficiency, we need to consider error-caused transition
between these two stationary states, i.e., full cooperation and full defection.
The transition from the latter to the former is possible only through state R,
which occurs with probability of $O(\epsilon^{n-1})$.
On the other hand, full cooperation can be made robust against every possible
type of $(n-1)$-bit error if $m=2n-1$.
To see the basic idea, let us suppose that memory is just long enough, and it
will immediately become clear what we mean by `enough'. If an initial state of
full cooperation is disturbed by error, everyone goes to either P (punish) or
A (accept), and what happens is the following: Those who have defected more will
accept punishment from the others, so the players' numbers of defections tend to
be equalized as time goes by. When everyone has finally defected the same
number of times, they all arrive at state A, where $c$ is the prescribed action.
Then, according to the first rule C, full cooperation will continue.
If we look at our example with Alice, Bob, and Charlie ($n=3$), \hl{we basically
mean that our strategy corrects every single-bit or double-bit
error when its memory length is given as $m=2n-1=5$: To see how this happens,
let us consider some possible types of error case by case.}
\begin{enumerate}
\item Single-bit error:
Imagine that a player, say, Alice, mistakenly defects from full cooperation at
$t=1$. She will have state A at $t=2$, while the others have state P, so their
payoffs should be equalized at $t=3$ as a result of punishment, by which mutual
cooperation is recovered. This scenario can be represented as follows:
\begin{equation}
\begin{matrix}
   & t = 1 & & t = 2 & & t = 3 \\
A: & cccc\underline{d} & \xrightarrow[\text{A}]{} & cccdc &
\xrightarrow[\text{A}]{} & ccdcc\\
B: & ccccc & \xrightarrow[\text{P}]{} & ccccd & \xrightarrow[\text{A}]{} & cccdc
\\
C: & ccccc & \xrightarrow[\text{P}]{} & ccccd & \xrightarrow[\text{A}]{} &
cccdc,
\end{matrix}
\label{eq:single}
\end{equation}
where the underline means a mistaken action, and the letter below each arrow
means which rule applies there. The important point is that a single-bit error
is corrected only in two rounds.
\item Double-bit error:
In this case, we have several possibilities. First, we consider two players'
simultaneous mistakes, which are corrected in a similar way to
Eq.~\eqref{eq:single}.
\begin{equation}
\begin{matrix}
   & t = 1 & & t = 2 & & t = 3 \\
A: & cccc\underline{d} & \xrightarrow[\text{A}]{} & cccdc &
\xrightarrow[\text{A}]{} & ccdcc\\
B: & cccc\underline{d} & \xrightarrow[\text{A}]{} & cccdc &
\xrightarrow[\text{A}]{} & ccdcc \\
C: & ccccc & \xrightarrow[\text{P}]{} & ccccd & \xrightarrow[\text{A}]{} &
cccdc.
\end{matrix}
\label{eq:double0}
\end{equation}
As another possibility, let us assume that Alice defects by mistake for two
successive rounds. It is a simple extension of the recovery pattern in
Eq.~\eqref{eq:single}:
\begin{equation}
\begin{matrix}
   & t = 1 & & t = 2 & & t = 3 & & t = 4\\
A: & cccc\underline{d} & \xrightarrow[\text{A}]{} & cccd\underline{d} &
\xrightarrow[\text{A}]{} & ccddc & \xrightarrow[\text{A}]{} & cddcc \\
B: & ccccc & \xrightarrow[\text{P}]{} & ccccd & \xrightarrow[\text{P}]{} & cccdd
& \xrightarrow[\text{A}]{} & ccddc \\
C: & ccccc & \xrightarrow[\text{P}]{} & ccccd & \xrightarrow[\text{P}]{} & cccdd
& \xrightarrow[\text{A}]{} & ccddc.
\end{matrix}
\label{eq:double1}
\end{equation}
It makes little difference whether error occurs to a single
player twice in a row or it does to one after another:
\begin{equation}
\begin{matrix}
   & t = 1 & & t = 2 & & t = 3 & & t = 4\\
A: & cccc\underline{d} & \xrightarrow[\text{A}]{} & cccdc &
\xrightarrow[\text{A}]{} & ccdcc & \xrightarrow[\text{A}]{} & cdccc \\
B: & ccccc & \xrightarrow[\text{P}]{} & cccc\underline{c} & \xrightarrow[\text{P}]{} & ccccd & \xrightarrow[\text{A}]{} & cccdc \\
C: & ccccc & \xrightarrow[\text{P}]{} & ccccd & \xrightarrow[\text{A}]{} & cccdc
& \xrightarrow[\text{A}]{} & ccdcc.
\end{matrix}
\label{eq:double2}
\end{equation}
The last possibility to consider is when error occurs again at the end of
Eq.~\eqref{eq:single}:
\begin{equation}
\begin{matrix}
   & t = 1 & & t = 2 & & t = 3 & & t = 4 & & t = 5\\
A: & cccc\underline{d} & \xrightarrow[\text{A}]{} & cccdc
& \xrightarrow[\text{A}]{} & ccdc\underline{d}
& \xrightarrow[\text{A}]{} & cdcdc
& \xrightarrow[\text{A}]{} & dcdcc\\
B: & ccccc & \xrightarrow[\text{P}]{} & ccccd
& \xrightarrow[\text{A}]{} & cccdc
& \xrightarrow[\text{P}]{} & ccdcd
& \xrightarrow[\text{A}]{} & cdccc \\
C: & ccccc & \xrightarrow[\text{P}]{} & ccccd
& \xrightarrow[\text{A}]{} & cccdc
& \xrightarrow[\text{P}]{} & ccdcd
& \xrightarrow[\text{A}]{} & cdccc,
\end{matrix}
\label{eq:double3}
\end{equation}
which needs additional two rounds to reach full cooperation at $t=5$. Among all
types of double-bit error, the last pattern and the like (i.e., error
occurs again when cooperation is about to be recovered) are the ones that
require the longest memory for full
recovery: If the distance between two errors is longer than two rounds,
they can be regarded as two single-bit errors, which are separately correctable
[Eq.~\eqref{eq:single}]. In general, if we have to correct $(n-1)$-bit error
that occurs every
other round, the memory length $m=2(n-1)+1$ is required in total, where the last
bit has been added to memorize the last round of full cooperation. \hl{It} is
also enough to correct simpler types of error as illustrated above.
To sum up, with $m=2n-1$, the transition probability from mutual
cooperation to defection can be suppressed down to $O(\epsilon^{n})$, whereas
the transition in the opposite direction through R has probability of
$O(\epsilon^{n-1})$. Therefore, the players form full cooperation in the limit
of $\epsilon \to 0$, fulfilling the efficiency criterion.
\end{enumerate}

\subsubsection*{Distinguishability}
The last criterion is distinguishability.
If the others are AllC players, our CAPRI-$n$ player will continue unilateral
defection when she defected $n$ consecutive times by error, as prescribed by
\hl{rule} I.
One can escape from such a state with probability of $O(\epsilon^n)$
due to the condition of $N_d < n$ \hl{in rule C}, so this
stationary state coexists with full cooperation in the limit of $\epsilon \to
0$.

\subsection*{Evolutionary simulation}
\label{appendix:method}

We consider a standard stochastic model proposed in~\cite{imhof2010stochastic},
where a well-mixed population of size $N$ evolves over time by an imitation
process.
A key assumption of this model is that the mutation rate is low so that
at most one mutant strategy can exist in the resident population. In other
words, the time that it takes to go extinct or occupy the whole population by
selection is assumed to be much shorter than the time scale of mutation.
Let us assume that
a mutant strategy $x$ is introduced to a population of strategy $y$.
The population dynamics is modeled by the frequency-dependent Moran process, in
which the fixation probability of the mutant is given in a closed from:
\begin{equation}\label{eq:fixation_prob}
    \phi_{xy} = \left( \sum_{i=0}^{N-1} \prod_{j=1}^{i} \Gamma_j \right)^{-1}
\end{equation}
with $\Gamma_j \equiv P_{j,j-1}/P_{j,j+1}$, where $P_{j,j\pm 1}$
denotes the probability that the number of mutants increases or decreases
from $j$ by one.

For $n=2$, the fixation probability is calculated in the following way:
Suppose that we have $j$ individuals of the mutant strategy and $N-j$
individuals of the resident strategy. If we randomly
choose a mutant and a resident individual, their average payoffs are obtained
as
\begin{equation}
\begin{cases}
s_x = \frac{1}{N-1}\left[(j-1)s_{xx} + (N-j)s_{xy}\right] \\
s_y = \frac{1}{N-1}\left[(N-j-1)s_{yy} + js_{yx}\right],
\end{cases}
\end{equation}
respectively, where $s_{\alpha \beta}$ is $\alpha$'s long-term payoff
against $\beta$. According to
the imitation process, $x$ can change to $y$ with probability $f_{x
\to y}$ defined as follows:
\begin{equation}\label{eq:imit_dynamics}
    f_{x \to y} = \frac{1}{1 + \exp{\left[ \sigma (s_x - s_y) \right]}},
\end{equation}
where $\sigma$ means the strength of selection. Then, we have
\begin{equation}\label{eq:Gamma}
\Gamma_j = \exp\left[\sigma(s_y - s_x)\right],
\end{equation}
and the fixation probability is calculated as
\begin{eqnarray}
\phi_{xy}^{-1} &=& \sum_{i=0}^{N-1}\prod_{j=1}^{i}
e^{\sigma \left[ (N-j-1)s_{yy} + js_{yx} - (j-1)s_{xx} -
(N-j)s_{xy} \right]/(N-1)} \\
  &=& \sum_{i=0}^{N-1}
  e^{\sigma i \left[(-i+2N-3)s_{yy} + (i+1)s_{yx} -
  (-i+2N-1)s_{xy} - (i-1)s_{xx} \right]/[2(N-1)]}.
\end{eqnarray}
If $y$ is a friendly rival, i.e. if $s_{yy} \geq s_{xx}$ and $s_{yy} \geq
s_{xy}$ in addition to $s_{yx} \geq s_{xy}$, Jensen's inequality shows that
$\phi_{xy} \leq 1/N$ for arbitrary $x$, indicating that $y$ has evolutionary
robustness for any $N$, $\rho$, and $\sigma$~\cite{murase2020five}.

For $n=3$, the fixation probability is calculated in a similar way. We randomly
pick up three players from a well-mixed population, and the respective average
payoffs of playing $x$ and $y$ can be written by using the binomial coefficients
as follows:
\begin{equation}
    \begin{cases}
s_x = \frac{1}{(N-1)(N-2)} \left[ \binom{j-1}{2} s_{xxx} + \binom{j-1}{1}\binom{N-j}{1} s_{xxy} + \binom{N-j}{2} s_{xyy} \right] \\
s_y = \frac{1}{(N-1)(N-2)} \left[ \binom{j}{2} s_{yxx} + \binom{N-j-1}{1}\binom{j}{1} s_{yyx} + \binom{N-j-1}{2} s_{yyy} \right],
\end{cases}
\end{equation}
where $s_{\alpha \beta \gamma}$ is $\alpha$'s long-term payoff against
$\beta$ and $\gamma$.
Plugging these expressions into Eqs.~\eqref{eq:fixation_prob} and
\eqref{eq:Gamma}, one can calculate the fixation probability $\phi_{xy}$ for the
three-person case as well.
Differently from the two-person case, however, friendly rivalry itself does not
necessarily guarantee evolutionary robustness if $n \ge 3$: Assume that a
friendly-rivalry strategy $y$
cannot distinguish a mutant $x$, whereas $x$ does distinguish $y$ when $x$ forms
the majority of the $n$-person game. If $n=3$, for example, it means that
$s_{yyx} = s_{xyy} = \rho$ whereas $s_{yxx} = s_{xxy} = 0$. Furthermore, if the
mutants are efficient among themselves, i.e., $s_{xxx} = \rho$, then its
fixation probability will be higher than $1/N$. As of now, we find no reason to
rule out the possibility of such a mutant.

We can interpret $\phi_{xy}$ as transition probability from $y$ to $x$ from the
viewpoint of the population.
From the stationary distribution of this Markovian dynamics,
we can thus calculate abundance of each available strategy in a numerically
exact manner~\cite{jeong2014optional,baek2016comparing}. For the sake of
simplicity, we use the donation game as a simplified form of
the PD game as well as its generalization to $n$ players in the
numerical calculation. That is, with the benefit of cooperation $b>1$, each
player can donate $b/(n-1)$ to each co-player at the unit cost, which
corresponds to $\rho = nb/[b+(n-1)]$ up to scaling.
In the next section, we will present numerical results obtained by using
OACIS~\cite{murase2018open}.

% For figure citations, please use "Fig" instead of "Figure".

% Place figure captions after the first paragraph in which they are cited.

% Results and Discussion can be combined.
\section*{Results}

\subsection*{Friendly rivalry}
To check the validity of our construction, we computationally examined the three criteria by
using graph-theoretic calculations used in~\cite{hougardy2010floyd,murase2018seven,murase2020five}.
For $n=2$, we directly confirmed that CAPRI-$2$ is indeed a successful strategy
satisfying all the three criteria. For $n=3$, we conducted mapping to an
automaton to obtain a simplified yet equivalent graph
representation~\cite{murase2019automata} to reduce computational complexity,
and our graph-theoretic calculation confirmed that the resulting automaton
indeed passed all the criteria.
For $n=4$, the required amount of calculation to directly check the criteria
was beyond our computational resources, so we employed a Monte Carlo method to
simulate the game. The Monte Carlo method was also used to double-check the
performance of CAPRI-$2$ and CAPRI-$3$.
See \nameref{S1_Appendix} for more discussion on computational details.

\begin{figure}
\begin{center}
\includegraphics[width=0.9\textwidth]{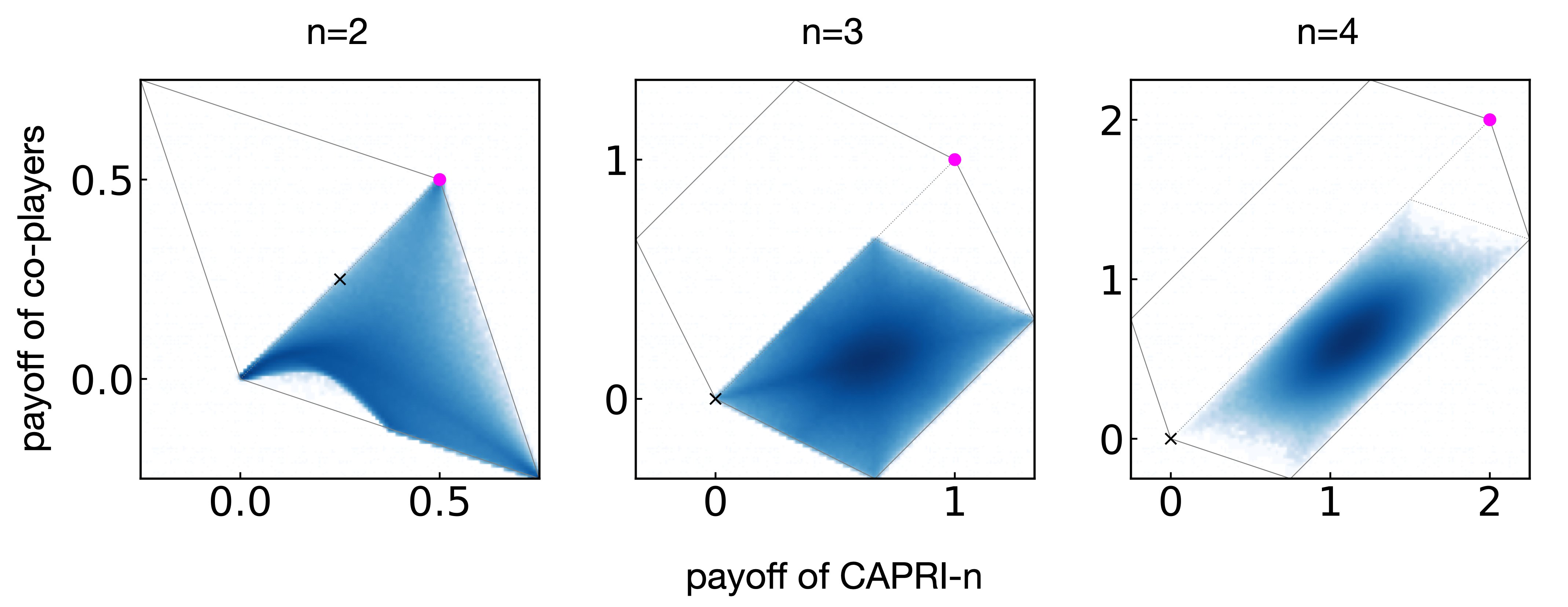}
\end{center}
\caption{
Distribution of long-term payoffs when a CAPRI-$n$ player meets co-players whose
$p_{\mu \nu}$'s are randomly sampled from the unit interval.
Darker shades toward blue indicate higher frequency of occurrence.
The multiplication factors for $n=2$, $3$, and $4$ are $1.5$, $2$, and $3$,
respectively, and the solid lines indicate the region of feasible payoffs.
In each case, the filled circle means the long-term payoffs when CAPRI-$n$ is
adopted by all the players, whereas the cross shows those of TFT players as a
reference point.
In each panel, we have drawn a dotted line along the diagonal as a simple check
for defensibility.
For $n=3$ or $4$, the parallelogram surrounding the blue area indicates
the set of feasible payoffs when the focal player is AllD, which indicates that
the behavior of CAPRI-$n$ is similar to AllD against most of the memory-one
players.
}
\label{fig:defensibility_distribution}
\end{figure}

The Monte Carlo calculation was performed as follows:
Let us denote a memory-one strategy as $(p_{cc}, p_{cd}, p_{dc}, p_{dd})$ where
$p_{\mu \nu}$ means the player's probability to cooperate when the player and
the co-player did $\mu$ and $\nu$, respectively, in the previous round. The
initial $\mu$ and $\nu$ can be omitted in the strategy description because they
are irrelevant to the long-term payoff as long as $\epsilon>0$.
\hl{Fig}~\ref{fig:defensibility_distribution} shows the distribution of payoffs
when Alice used CAPRI-$n$ whereas each of her co-players' strategies was
composed of four $p_{\mu \nu}$'s randomly sampled from the unit interval.
The co-players' payoffs never exceeded Alice's, as required by defensibility.

We also calculated the probability of full cooperation
for $n=2$, $3$ and $4$ when CAPRI-$n$ was adopted by all the players in order to
check efficiency.
By using linear-algebraic~\cite{yi2017combination,murase2018seven} or Monte Carlo calculation,
with $\epsilon = 10^{-4}$, we obtained $0.999$, $0.997$, $0.978$ for $n=2$, $3$,
and $4$, respectively, which supports the conclusion that they all satisfy the
efficiency criterion.

\subsection*{Evolutionary performance}

\begin{figure}
\begin{center}
\includegraphics[width=0.9\textwidth]{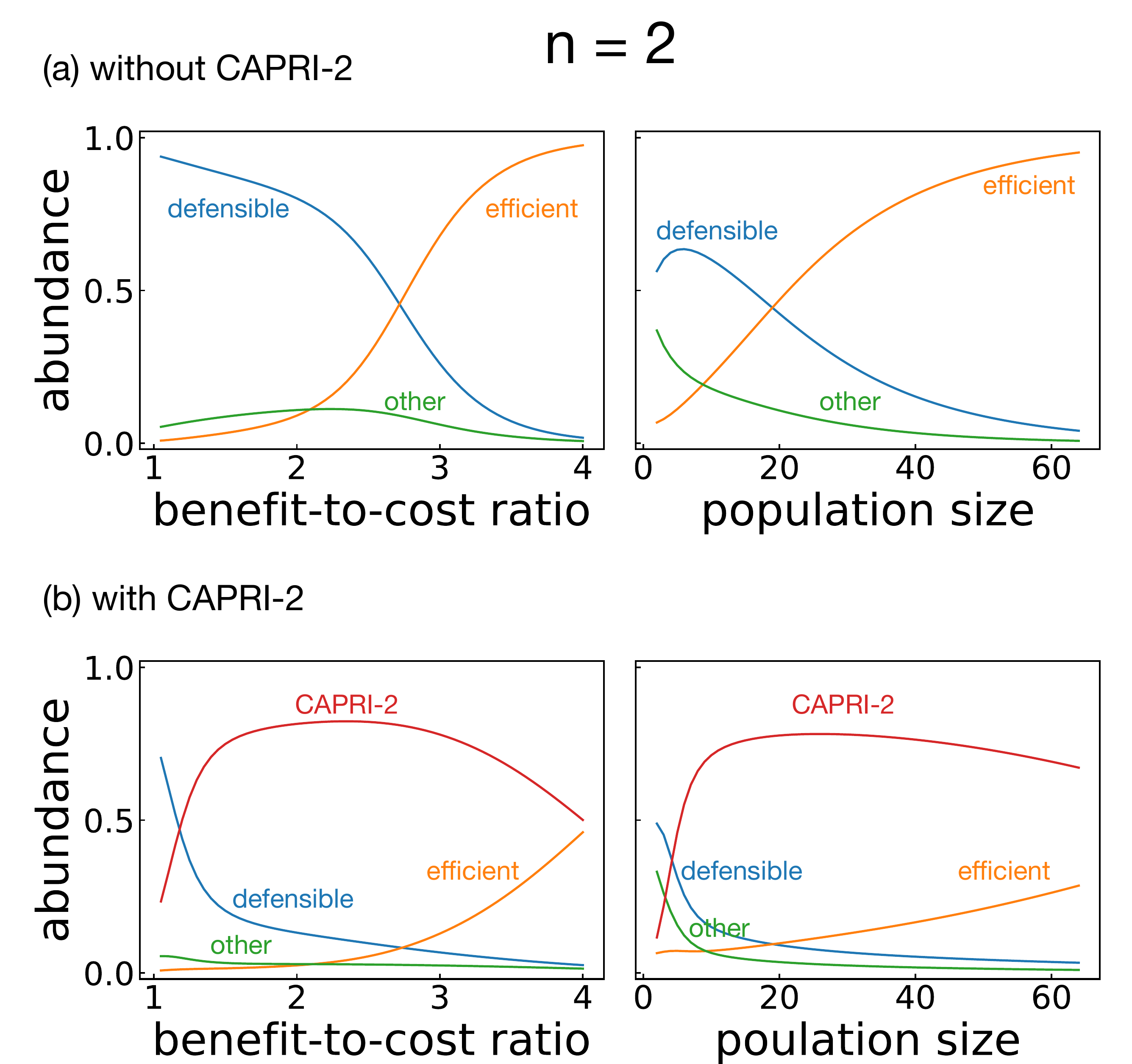}
\end{center}
\caption{Abundance of strategies for $n=2$ as the benefit-to-cost ratio $b$ and
the population size $N$ vary. The default values were $b=3$ and $N=30$ unless
otherwise specified. The strength of selection and the error probability
were set to be $\sigma=1$ and $\epsilon = 10^{-4}$, respectively.
\hl{(A)} Simulation result with $16$ memory-one deterministic strategies,
classified into three categories, i.e., efficient, defensible, and the
other strategies.
\hl{(B)} Effect of CAPRI-$2$ when it was added to the available set of strategies.
}
\label{fig:evo_n2}
\end{figure}

\begin{figure}
\begin{center}
\includegraphics[width=0.9\textwidth]{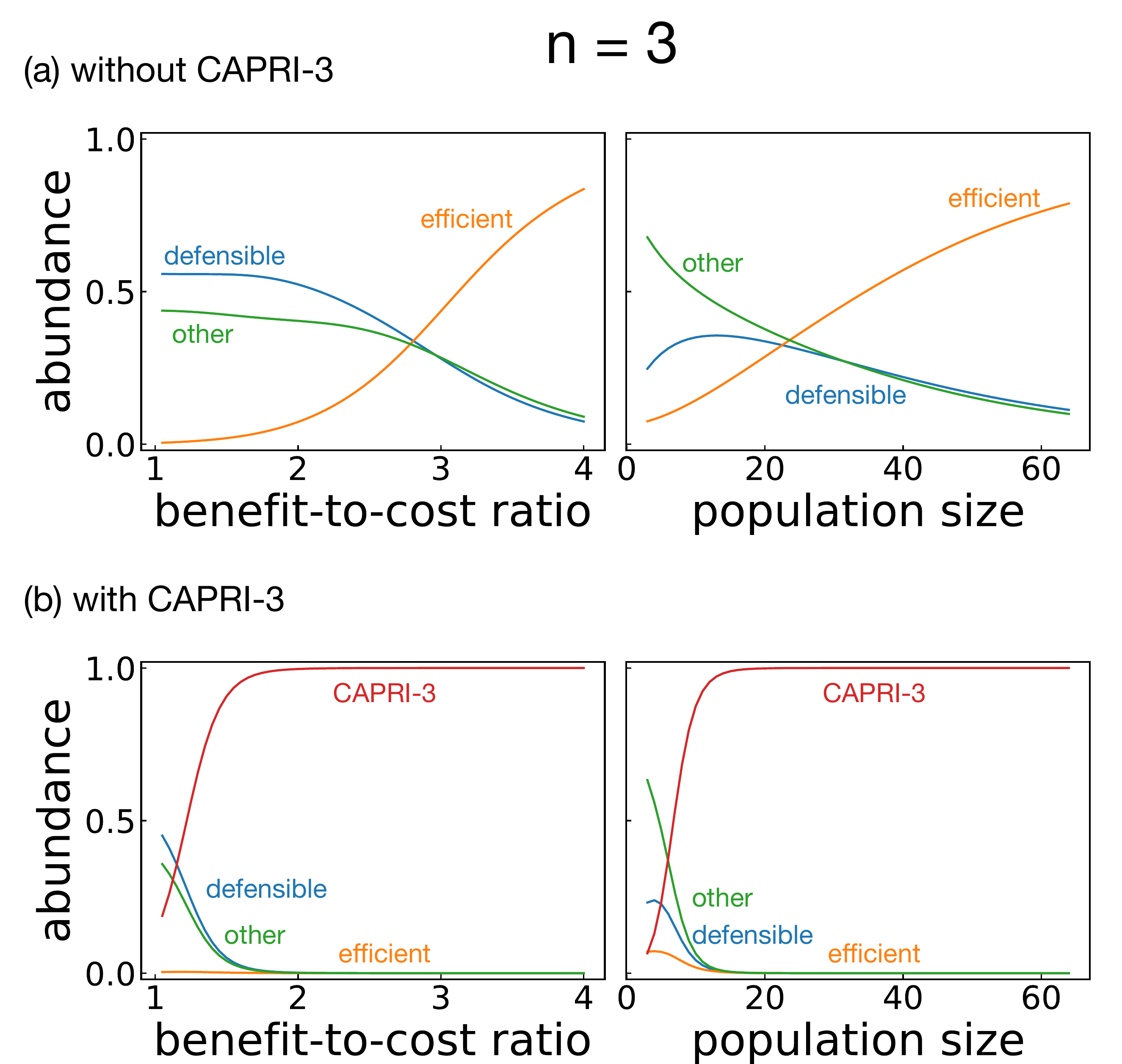}
\end{center}
\caption{Abundance of strategies for $n=3$ as the benefit-to-cost ratio $b$ and
the population size $N$ vary. The default values were $b=3$ and $N=30$ unless
otherwise specified. The strength of selection and the error probability
were set to be $\sigma=1$ and $\epsilon = 10^{-4}$, respectively.
\hl{(A)} Simulation result with $64$ memory-one deterministic strategies,
classified into three categories, i.e., efficient, defensible, and the
other strategies.
\hl{(B)} Effect of CAPRI-$3$ when it was added to the available set of strategies.
}
\label{fig:evo_n3}
\end{figure}

Before checking the evolutionary performance of our proposed strategy,
we conducted simulations without CAPRI-$n$ for comparison.
\hl{Figs~\ref{fig:evo_n2}A and~\ref{fig:evo_n3}A} show the results when
the strategies were sampled from deterministic memory-one for $n=2$ and $3$.
When $b$ was low and/or $N$ was small, defensible strategies such as AllD tended
to be favored by selection, and the resulting cooperation level was low.
On the other hand, when $b$ or $N$ was large, efficient strategies
were favored, and they achieved a high level of cooperation. The reason is that
cooperative strategies maintained high payoffs by interacting with many other
cooperators even if they were exploited by a small number of aggressive mutants.

When CAPRI-$n$ was introduced, it occupied a large amount of the population as
shown in \hl{Figs~\ref{fig:evo_n2}B and~\ref{fig:evo_n3}B}. Whereas each
memory-one strategy flourished depending on the environmental parameters $b$ and
$N$, CAPRI-$n$ was found abundant in the entire parameter region.
In particular, it is striking that CAPRI-$3$ overwhelms all the other
strategies in the three-person PG game for any moderate sizes of $b$ and
$N$ (Fig~\ref{fig:evo_n3}B).

It is nevertheless worth pointing out that
CAPRI-$2$ gave more and more room to efficient strategies in
the iterated PD game as $b$ or $N$ increases (\hl{Fig}~\ref{fig:evo_n2}B), and
this is due to neutral drift:
Although CAPRI-$2$ earns a strictly higher long-term payoff
than AllC$=(1,1,1,1)$ and Win-Stay-Lose-Shift (WSLS) $=(1,0,0,1)$, it does not
with respect to $(1,1,1,0)$, which can, in turn, be invaded by WSLS. For this
reason, WSLS can become abundant in the presence of $(1,1,1,0)$ when the
environmental conditions are favorable.

\begin{figure}
\begin{center}
\includegraphics[width=0.9\textwidth]{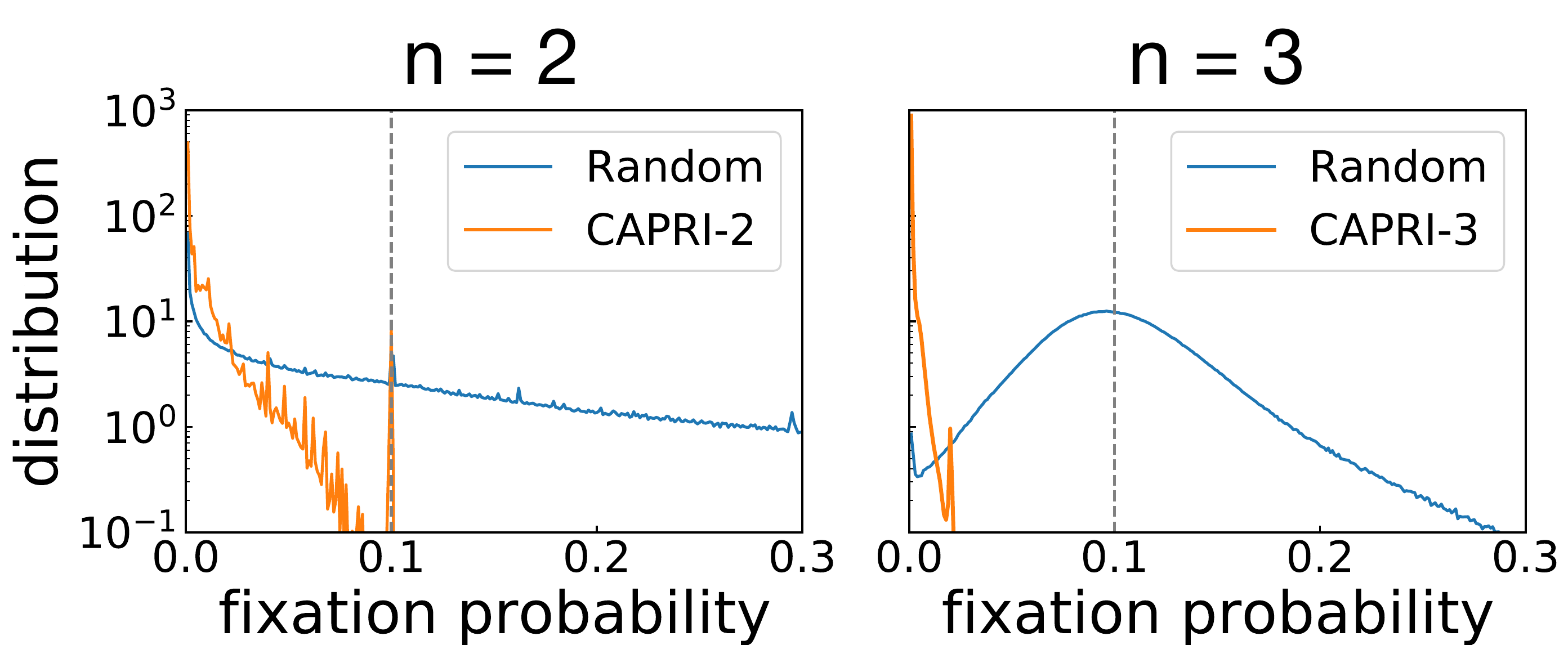}
\end{center}
\caption{
Normalized distribution of fixation probabilities of mutants, which were
randomly sampled from the set of deterministic strategies with the same
memory length as CAPRI-$n$'s.
When we simulated the two-person game with taking CAPRI-$2$ as the resident
strategy, $10^9$ mutants were sampled. In case of the three-person game in which
CAPRI-$3$ was the resident strategy, the number of sampled mutants was $5\times
10^6$. In either case, no mutant had higher fixation probability than $1/N$ (the
vertical dashed line).
On the other hand, when the resident was randomly drawn from the same strategy
set, mutants frequently achieved fixation with probability higher than $1/N$.
For each random sample of the resident strategy, $10^2$ mutants were tested, and
this process was repeated $10^7$ and $10^5$ times when $n=2$ and $3$,
respectively.
Throughout this calculation, we used $N=10$ as the population size,
$\epsilon=10^{-4}$ as the error probability, $\sigma=1$ as the selection
strength, and $b=2$ as the benefit of cooperation.
}
\label{fig:rho_distribution}
\end{figure}

We also tested the performance of CAPRI-$n$ against strategies with the same
memory length. An obvious problem is the huge number of possible strategies:
Provided that $m=2n-1$, the number amounts to $2^{2^{nm}} \sim 10^{19}$ for
$n=2$, which grows to $10^{9864}$ for $n=3$.
As an alternative to exhaustive enumeration,
we calculated fixation probabilities of mutants that are randomly sampled
from the set of deterministic strategies with $m=2n-1$, with taking CAPRI-$n$ as
the resident strategy. The numbers of sampled mutants were $10^9$ and $5 \times
10^6$ for
$n=2$ and $3$, respectively. As shown in \hl{Fig}~\ref{fig:rho_distribution}, none
of them had fixation probability greater than $1/N$, and the
tendency was more pronounced in the three-person game than in the two-person
case. For comparison, we also tested resident strategies drawn randomly from the
same strategy set with $m=2n-1$, and a significant fraction of mutants succeeded
in fixation with probability higher than $1/N$. Although we have no proof for
evolutionary robustness for $n \ge 3$, the numerical result suggested that it
would be extremely unlikely to see CAPRI-$n$ invaded by random mutants even if
they had the same memory length.

% Place tables after the first paragraph in which they are cited.
%\begin{table}[!ht]
%\begin{adjustwidth}{-2.25in}{0in} % Comment out/remove adjustwidth environment if table fits in text column.
%\centering
%\caption{
%{\bf Table caption Nulla mi mi, venenatis sed ipsum varius, volutpat euismod diam.}}
%\begin{tabular}{|l+l|l|l|l|l|l|l|}
%\hline
%\multicolumn{4}{|l|}{\bf Heading1} & \multicolumn{4}{|l|}{\bf Heading2}\\ \thickhline
%$cell1 row1$ & cell2 row 1 & cell3 row 1 & cell4 row 1 & cell5 row 1 & cell6 row 1 & cell7 row 1 & cell8 row 1\\ \hline
%$cell1 row2$ & cell2 row 2 & cell3 row 2 & cell4 row 2 & cell5 row 2 & cell6 row 2 & cell7 row 2 & cell8 row 2\\ \hline
%$cell1 row3$ & cell2 row 3 & cell3 row 3 & cell4 row 3 & cell5 row 3 & cell6 row 3 & cell7 row 3 & cell8 row 3\\ \hline
%\end{tabular}
%\begin{flushleft} Table notes Phasellus venenatis, tortor nec vestibulum mattis, massa tortor interdum felis, nec pellentesque metus tortor nec nisl. Ut ornare mauris tellus, vel dapibus arcu suscipit sed.
%\end{flushleft}
%\label{table1}
%\end{adjustwidth}
%\end{table}

%PLOS does not support heading levels beyond the 3rd (no 4th level headings).
\section*{Discussion}

In summary, we have constructed a friendly-rivalry strategy for the iterated
$n$-person PG game. It maintains a cooperative Nash equilibrium in the
presence of implementation error with probability $\epsilon \ll 1$, and
it shows excellent evolutionary performance
regardless of the environmental conditions such as the
population size and the strength of selection.
In this sense, the $n$-person social dilemma is solved.
The strategy requires memory of the previous $m=2n-1$
rounds and consists of the following five rules: Cooperate if
everyone did, accept punishment for your own mistake, punish others' defection,
recover cooperation if you find a chance, and defect in all the other
circumstances.

Although we have considered only
implementation error, perception error can also be corrected if it occurs with
sufficiently low probability: The disagreement between the players' history
profiles due to the perception error will soon be removed at full
defection, and the players will escape from mutual defection
with probability of $O(\epsilon^n)$. Unless another perception error perturbs
this
process, the players will eventually arrive at full cooperation, overcoming the
perception error.

Another important solution concept to the $n$-person dilemma can be derived from
a different set of criteria: By requiring mutual cooperation, error correction,
and retaliation with a time scale of $k$ rounds, one can characterize the
all-or-none (AON-$k$) strategy, which is defined as prescribing $c$ only when
everyone cooperated or no one did in each of the previous $k$
rounds~\cite{hauert1997effects,lindgren1997evolutionary,hilbe2017memory}.
For example, WSLS$=(1,0,0,1)$ is equivalent to AON-$1$.
For each $k$, one can find a threshold of the multiplication factor above which
AON-$k$ constitutes a subgame-perfect equilibrium~\cite{hilbe2017memory}.
AON-$k$ performs well in evolutionary simulation because it prescribes $d$ as
the default action, just as CAPRI-$n$ does in state I, unless the players have
synchronized their behavior over the previous $k$ rounds. As a result, it earns
a strictly higher payoff against a broad range of strategies.

In general, CAPRI-$n$ with $m=2n-1$ can repeatedly exploit the other co-players
playing AON-$k$ if $k < m-1$, which means that an AON-$k$ population can readily
be invaded by CAPRI-$n$ unless $k$ is large enough.
Considering the condition for AON-$k$ to be subgame perfect, one could speculate
that AON with small $k$ can be abundant in an environment with a high
multiplication factor. However, our finding implies that such a simple solution
may not be sustained when CAPRI-$n$ is available.
This is especially crucial when population size is not large enough because
AON-$k$ lacks defensibility.
Still, AON-$k$ remains as a strong competitor to CAPRI-$n$ in evolutionary
simulation: For example, although WSLS earns a strictly less payoff against
CAPRI-$2$, it circumvented the difficulty of fixation with the aid of a third
strategy $(1,1,1,0)$.

We have assumed the small-$\epsilon$ limit, but an important question is how
the performance of CAPRI-$n$ will be affected if $\epsilon$ takes a finite
value. One possibility is that it could set a limit on the maximum number of
players in regard to the efficiency criterion: The
transition probability from full defection to cooperation is given as $n
\epsilon^{n-1}$ by construction, where the prefactor $n$ originates from the
number of possible ways to choose $(n-1)$ players who will cooperate by mistake.
The probability of transition in the opposite direction is of
$O(\epsilon^n)$ at most, but it is reasonable to guess that it also has a
prefactor as an increasing function of $n$. If it can be approximated as $n^\tau
\epsilon^{n}$ with $\tau>1$, for example, efficiency criterion should require $n
\lesssim \left( 1/\epsilon \right)^{1/(\tau-1)}$. To achieve cooperation among a
large number of players with finite error probability, therefore, we may have to
revise the rules so as to adjust the prefactors.

From a practical point of view, it is worth noting that the five rules of
CAPRI-$n$ mostly refer to two factors: One is the players' last action at $t-1$,
and the other is the differences in the players' respective numbers of
defections over the previous $m$ rounds.  In other words, exact details of the
history profile are irrelevant, and this point greatly reduces the cognitive
burden to play this strategy. In fact, according to a recent experiment, people
assign reputation to their co-players based mainly on their last action and their average numbers of defection~\cite{cuesta2015reputation}.
This could explain the reason that such a delicate relationship called friendly
rivalry can develop spontaneously and unwittingly among a group of people.
How to keep such a relation healthy and productive has so far been acquired as
tacit knowledge surrounded by anecdotes and experiences, and CAPRI-$n$ expresses
its essential how-tos in a form of explicit knowledge which can be designed,
analyzed, and transmitted systematically.

\section*{Acknowledgments}
%Y.M. acknowledges support from MEXT as
%``Exploratory Challenges on Post-K computer (Studies of multi-level
%spatiotemporal simulation of socioeconomic phenomena)'' and from Japan Society
%for the Promotion of Science (JSPS) (JSPS KAKENHI; Grant no. 18H03621). S.K.B.
%acknowledges support by Basic Science Research Program through the National
%Research Foundation of Korea (NRF) funded by the Ministry of Education
%(NRF-2020R1I1A2071670).
%This research used computational resources of the K computer provided by the
%RIKEN Center for Computational Science through the HPCI System Research project
%(Project ID:hp160264).
%Y.M. acknowledges support from Japan Society for the Promotion of Science (JSPS) (JSPS KAKENHI; Grant no. 18H03621). S.K.B. acknowledges support by Basic Science Research Program through the National Research Foundation of Korea (NRF) funded by the Ministry of Education (NRF-2020R1I1A2071670).
Part of the results is obtained by using the Fugaku computer at RIKEN Center for Computational Science (Proposal number ra000002).
\hl{We appreciate the APCTP for its hospitality during completion of this work.}

%\nolinenumbers

% Either type in your references using
% \begin{thebibliography}{}
% Text
% \end{thebibliography}
%
% or
%
% Compile your BiBTeX database using our plos2015.bst
% style file and paste the contents of your .bbl file
% here. See http://journals.plos.org/plosone/s/latex for
%\bibliography{main}

%For more information, see \nameref{S1_Appendix}.

\section*{Supporting information}
\setcounter{equation}{0}
\renewcommand{\theequation}{S\arabic{equation}}

\paragraph*{S1 Appendix.}
\label{S1_Appendix}
{\bf Computational check for efficiency, defensibility, and distinguishability}
\setcounter{table}{0}
\setcounter{equation}{0}
\renewcommand{\theequation}{S\arabic{equation}}

\subsection*{Efficiency}

The most direct method for checking efficiency is to construct a
Markovian transition matrix. For example, if we check efficiency of WSLS,
the matrix can be written as
\begin{eqnarray}
&\begin{array}{cccc}
\multicolumn{4}{c}{h_{t-1}}\\
(c,c) & ~~~~(c,d) & ~~~~(d,c) & ~~~~(d,d)~~~~
\end{array}\nonumber\\
\begin{array}{cc}
\multirow{4}{*}{\smash{\rotatebox[origin=c]{90}{$h_t$}}}&(c,c)\\
&(c,d)\\
&(d,c)\\
&(d,d)
\end{array}
&\left[
\begin{array}{cccc}
(1-\epsilon)^2 & \epsilon^2 & \epsilon^2 & (1-\epsilon)^2 \\
\epsilon(1-\epsilon) & \epsilon(1-\epsilon) & \epsilon(1-\epsilon) & \epsilon(1-\epsilon) \\
\epsilon(1-\epsilon) & \epsilon(1-\epsilon) & \epsilon(1-\epsilon) & \epsilon(1-\epsilon) \\
\epsilon^2 & (1-\epsilon)^2 & (1-\epsilon)^2 & \epsilon^2
\end{array}
\right],
\label{eq:mat}
\end{eqnarray}
where each element is the conditional probability to observe $h_t = (A_t, B_t)$
for given $h_{t-1} = (A_{t-1}, B_{t-1})$. The stationary probability
distribution of the above process can be calculated explicitly if $m$ is
small~\cite{you2017chaos}, but it is often easier to obtain a numerical
solution, e.g., by using the QR algorithm~\cite{newman2013computational}.
If the stationary probability of full cooperation approaches $100\%$ as
$\epsilon$ decreases, the strategy is judged as efficient.

Note that the above direct method calculates the effect of every possible type
of error all at once. A quicker
way is to look at the connection structure of history
profiles~\cite{murase2020five}, and let us only outline the idea briefly:
In the above example of WSLS, suppose that $\epsilon=0$ as the zeroth-order
approximation of the actual process. It is then immediately clear that from
every initial condition the players should eventually end up with $(c,c)$ and
stay there. Once this is the case, we can prove that the existence of $\epsilon
\ll 1$ perturbs the stationary distribution only to a marginal degree so that we
will actually get the same answer as $\epsilon \to 0^+$. Or, if
the zeroth-order approximation fails to give such a definite answer, then we
have to modify the connection structure to the first order by including
transitions with probability of $O(\epsilon)$. This process is repeated with
increasing the order of $\epsilon$ one by one, until we reach the answer. One
may refer to Ref.~\cite{murase2020five} for the details.

\subsection*{Defensibility}
When a focal player's strategy is
given, his or her action is determined at every history profile, whereas
the other $(n-1)$ co-players' are not. Therefore, each history profile can lead
to $2^{n-1}$ possible history profiles in the next round.
If each history profile is regarded as
a node, we thus obtain a connection structure in which every node has $2^{n-1}$
outward links. At every node, we can see the focal player's instantaneous payoff
based on the actions taken in round $t$. The point is that we are concerned
about the focal player's long-term average payoff, to which only results from
cyclic loops can contribute. The basic idea for checking defensibility is thus
to find all the loops in the connection structure: If a loop puts the focal
player at a payoff disadvantage compared with the co-players, let us call it a
negative loop. If the connection structure contains no such negative loops, the
corresponding strategy is defensible because it is impossible to make the
player's long-term average payoff lower than the co-players' no matter what they
choose. To detect all the negative loops, we have used the Floyd-Warshall
algorithm~\cite{hougardy2010floyd,murase2018seven,murase2020five}.

\subsection*{Distinguishability}
The distinguishability criterion can be checked in a similar way to the one for
the efficiency criterion because it is essentially determined from the
stationary distribution. For example, we can construct a transition matrix
between a given strategy and AllC as in Eq.~\eqref{eq:mat}. For the strategy
to be distinguishable, the stationary distribution must keep a finite amount of
probability to observe defection against AllC as $\epsilon \to 0^+$.

%%
%%% Include only the SI item label in the paragraph heading. Use the \nameref{label} command to cite SI items in the text.
\paragraph*{S1 Table.}
\label{S1_Table}
{\bf Summary of mathematical symbols used in this work.}

\setcounter{table}{0}
\renewcommand{\thetable}{S\arabic{table}}
%
%% Include only the SI item label in the paragraph heading. Use the \nameref{label} command to cite SI items in the text.
\begin{table}[htb]
\begin{center}
%\caption{
%{\color{black}
%{\bf Summary of mathematical symbols used in this work.}
%}
%}
\label{tab:notations}
\begin{tabular}{|c|l|}
\hline
$n$ & number of players \\
$m$ & memory length \\
$t$ & time in units of rounds\\
$\rho$ & multiplication factor of the public-goods game \\
$\epsilon$ & probability of implementation error \\
$n_c$ & number of cooperators \\
$X_t$ & player X's action at $t$\\
$h_t$ & history profile composed of the players' actions over the past $m$
rounds\\
$\pi_X^{(t)}$ & player $X$'s instantaneous payoff in round $t$ \\
$\Pi_X$ & player $X$'s long-term average payoff \\
$\Delta_A^{\tau_1,\tau_2}$ & Alice's number of defections during a time interval $[ \tau_1, \tau_2 ]$ \\
$t^{\ast}$ & the last round of full cooperation\\
$N_d$ & maximum difference in numbers of defections in memory \\
$N$ & population size in an evolutionary game \\
$P_{j,j\pm 1} $ & probability that the number of mutants increases or decreases
from $j$ by one\\
$\Gamma_j$ & $P_{j,j-1} / P_{j,j+1}$\\
$f_{x\to y}$ & probability that a player changes the strategy from $x$ to $y$\\
$\sigma$ & strength of selection\\
$\phi_{xy}$ & fixation probability of mutant $x$ in a resident $y$-population \\
$s_{\alpha\beta}$ & $\alpha$'s long-term payoff against $\beta$ in the
two-person PG game\\
$s_{\alpha\beta\gamma}$ & $\alpha$'s long-term payoff against $\beta$ and
$\gamma$ in the three-person PG game\\
$s_x$ & average payoff of $x$ \\
$b$ & benefit of cooperation in the donation game\\
$p_{\mu\nu}$ & probability to cooperate when two players did $\mu$ and $\nu$,
respectively\\
\hline
\end{tabular}
\end{center}
\end{table}

\end{document}